\documentclass[aps,prl,twocolumn,showpacs,floatfix]{revtex4}
\usepackage{amsmath}
\usepackage{epsfig}

\begin{document}

\title{Optimizing the Signal to Noise Ratio of a Beam Deflection Measurement with Interferometric Weak Values} 
\author{David J. Starling, P. Ben Dixon, Andrew N. Jordan and John C. Howell}
\affiliation{Department of Physics and Astronomy, University of Rochester, Rochester, New York 14627, USA}

\begin{abstract}
The amplification obtained using weak values is quantified through a detailed investigation of the signal to noise ratio for an optical beam deflection measurement. We show that for a given deflection, input power and beam radius, the use of interferometric weak values allows one to obtain the optimum signal to noise ratio using a coherent beam. This method has the advantage of reduced technical noise and allows for the use of detectors with a low saturation intensity. We report on an experiment which improves the signal to noise ratio for a beam deflection measurement by a factor of 54 when compared to a measurement using the same beam size and a quantum limited detector.
\end{abstract}


\maketitle

The ultimate limit of the sensitivity of a beam deflection measurement is of great interest in physics. The signal to noise ratio (SNR) of such measurements is limited by the power fluctuations of coherent light sources such as a laser, providing a theoretical bound known as the standard quantum limit \cite{Braginskii1975}. It was found that interferometric measurements of longitudinal displacements and split-detection of transverse deflections have essentially the same ultimate sensitivity \cite{Putman1992}. In this Rapid Communication we consider a beam deflection measurement technique that combines interferometry with split detection.  The technique makes use of weak values and results in the same ultimate sensitivity, but with a number of advantages for precision measurement science.

Weak values were introduced in 1988 by Aharonov \textit{et al.}\ \cite{Aharonov1988}. They claimed that the measurement of a component of the spin of a spin-1/2 particle can turn out to be 100, far outside the eigenvalue range of the measurement operator. More recently, the phenomenon known as weak values has been explored in the field of quantum optics \cite{Ritchie1990,Pryde2005,Hosten2008,Dixon2009} and solid state physics \cite{Williams2008,Romito2008}. Typically, a weak value experiment goes as follows: (1) pre-selection of an initial quantum state; (2) a weak interaction that couples a two-state observable (the \textit{system}) with a continuous variable (the \textit{meter}); and (3) post-selection on a state nearly orthogonal to the pre-selected \textit{system} state. The \textit{meter} variable is the measured, amplified parameter. This scheme throws away most of the data with the post-selection and yet, as we will show, the amplification of the measured parameter outweighs this effect.

In an interferometric weak value setup measuring beam deflection [caused by a piezo-actuated (PA) mirror], Dixon \textit{et al.}\ \cite{Dixon2009,Howell2009} used the which-path degree of freedom (the \textit{system} observable) of a Sagnac interferometer coupled with the transverse degree of freedom (the \textit{meter} variable) of a laser beam (see Fig. 1). With this method, they measured the angular deflection of a beam down to 400 femtoradians.

Standard techniques to optimize the SNR of a beam deflection measurement include focusing the beam onto a split detector or focusing the beam onto the source of the deflection. The improvement of the SNR is of great interest in not only deflection and interferometric phase measurements but also in spectroscopy and metrology \cite{Bollinger1996,Huelga1997}, anemometry \cite{Li1997}, positioning \cite{Giovannetti2001}, micro-cantilever cooling \cite{Kleckner2006}, and atomic force microscopy  \cite{Meyer1988,Rugar1989}. In particular, atomic force microscopes are capable of reaching atomic scale resolution using either a direct beam deflection measurement \cite{Meyer1988} or a fiber interferometric method \cite{Rugar1989}. We show that for any given beam radius, interferometric weak value amplification (WVA) can improve (or, at least match) the SNR of such beam deflection measurements. It has also been pointed out by Hosten and Kwiat that WVA reduces technical noise, which combined with our result provides a powerful technique \cite{Hosten2008}. 

\begin{figure}
\centerline{\includegraphics[scale=0.75]{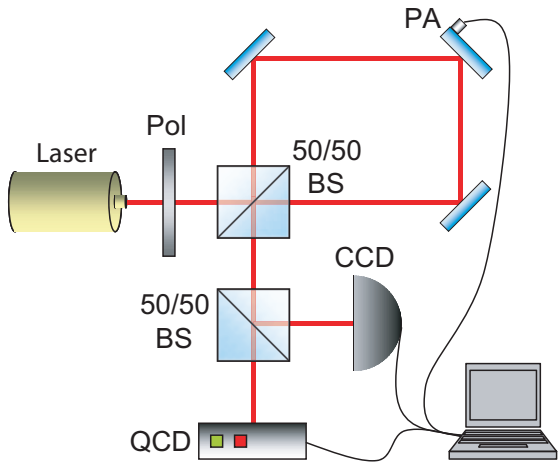}}
\caption{(Color online) A fiber coupled laser beam is launched into free space before passing through a polarizer, producing a horizontally polarized single mode Gaussian beam. The laser enters the input port of a Sagnac interferometer via a 50/50 beamsplitter (BS). The light is divided equally and travels through the interferometer clockwise and counterclockwise, encountering three mirrors before returning to the BS. The piezo-actuated mirror (PA), positioned symmetrically in the interferometer, causes a slight opposite deflection for the two different paths, altering the interference at the BS. The dark port is monitored with both a CCD camera and a quadrant cell detector (QCD) positioned at equal lengths from the second BS. The CCD is used only to verify the mode quality of the dark port.}
\end{figure}

The analogy between interferometry and beam deflection described in a paper by Barnett \textit{et al.}\ \cite{Barnett2003} allows one to predict the SNR for a deflection of an arbitrary optical beam (e.g.\ coherent or squeezed). For a coherent beam with a horizontal Gaussian intensity profile at the detector of
\begin{equation}
I(x) = \frac{1}{\sqrt{2\pi}\sigma}e^{-x^2/2\sigma^2},
\label{intensity}
\end{equation}
they show that the SNR is given by 
\begin{equation}
\mathcal{R} = \sqrt{\frac{2}{\pi}}\frac{\sqrt{N}d}{\sigma},
\label{SNR}
\end{equation}
where $N$ is the total number of photons incident on the detector, $d$ is the transverse deflection, and $\sigma$ is the beam radius defined in Eq.\ (\ref{intensity}). Equation (\ref{SNR}) represents the ultimate limit of the SNR for position detection with a coherent Gaussian beam. 

We now incorporate weak values by describing the amplification of a deflection at a split detector as a multiplicative factor $\cal A$. Thus, $d_a = {\cal A} d$ is the amplified deflection caused by the weak value. Also, the post-selection probability $P_{ps}$ modifies the number of photons incident on the detector such that $N_a = P_{ps} N$. The beam radius is not altered. Dixon \textit{et al.}\ showed that for a collimated Gaussian beam passing through a Sagnac interferometer (see Fig. 1) the WVA factor and the post-selection probability are given by 
\begin{eqnarray}
{\cal A} = \frac{2k_{0}\sigma^2}{l_{md}} \cot({\phi/2}),\,\, P_{ps} = \sin^2(\phi/2),
\label{AandP}
\end{eqnarray}
where $l_{md}$ is the distance from the piezo-actuated mirror to the detector, $k_0$ is the wave number of the light and $\phi$ is the relative phase of the two paths in the interferometer. 

Using Eqs. (\ref{AandP}) and making the substitutions $d \rightarrow {\cal A} d$ and $N \rightarrow P_{ps} N$ into Eq.\ (\ref{SNR}), we find the weak value amplified SNR,
\begin{equation}
\mathcal{R}_A =  \alpha \, \mathcal{R},
\label{SNRA}
\end{equation}
where $\alpha = 2k_0\sigma^2 \cos(\phi/2)/l_{md}$. For a typical value of $\phi$ we note that $\cos(\phi/2) \approx 1$. 

Dixon \textit{et al.}\ extend their analysis by inserting a negative focal length lens before the interferometer, creating a diverging beam. This modifies the WVA such that the new SNR is given by
 \begin{eqnarray}
 \mathcal{R}'_A &=& \alpha \, \mathcal{R} \frac{l_{lm}+al_{md}/\sigma}{l_{lm}+l_{md}} = C \left(\sigma + a\frac{l_{md}}{l_{lm}}\right),
\label{SNRA2}
\end{eqnarray}
where $C = \sqrt{(8 N)/\pi}(k_0 l_{lm} d \cos(\phi/2))/(l_{md}(l_{lm}+l_{md}))$ and $a$ is the radius of the beam at the lens which is a distance $l_{lm}$ from the piezo-actuated mirror. It is interesting to note that the dependence of the SNR is proportional to the beam radius at the detector in the amplified case [Eq.\ (\ref{SNRA2})] but inversely proportional when there is no amplification [Eq.\ (\ref{SNR})].

Equations (\ref{SNRA}) and (\ref{SNRA2}) are the main theoretical results of this paper. We see that it is possible to greatly improve the SNR in a deflection measurement with experimentally realizable parameters. Typical values for the experiment to follow are $\phi/2 = 25^\circ$, $\sigma = 1.7$ mm, $l_{md} = 14$ cm and $k_0 = 8 \times 10^6 \mbox{ m}^{-1}$ such that the expected SNR amplification is $\alpha \approx 300$.

We notice that for small $\phi$, the value of $\alpha$ is the ratio of the SNR for a beam deflection measurement in the far-field and the near-field. The far-field measurement can be obtained at the focal plane of a lens. This is recognized as a typical method to reach the ultimate precision for a beam deflection measurement \cite{Putman1992}. Consider a collimated Gaussian beam with a large beam radius $\sigma$ which acquires a transverse momentum shift $k$ given by a movable mirror. The beam then passes through a lens with focal length $f$ followed by a split detector. The total distance from the source of the deflection to the detector is $l_{md}$, and the detector is at the focal plane of the lens. This results in a new deflection $d'=fk/k_0$ and a new beam radius $\sigma'=f/{2k_0\sigma}$ at the detector. Making the substitutions $d \rightarrow d'$ and $\sigma \rightarrow \sigma'$ into Eq.\ (\ref{SNR}), we see that when the beam is focused onto a split detector the SNR is amplified:  
\begin{equation}
\mathcal{R}_{f} = \alpha_f \, \mathcal{R},
\label{SNR'}
\end{equation}
where $\alpha_f = 2k_0\sigma^2/l_{md}$ is the improvement in the SNR relative to the case with no lens [i.e.\ Eq.\ (\ref{SNR})]. Yet this is identical to the improvement obtained using interferometric weak values, up to a factor of $\cos(\phi/2) \approx 1$ for small $\phi$. Thus we see that the improvement factors are equal using either WVA or a lens focusing the beam onto a split detector, resulting in the same ultimate limit of precision. However, WVA has three important advantages: the reduction in technical noise, the ability to use a large beam radius and lower intensity at the detector due to the post selection probability $P_{ps} = \sin^2(\phi/2)$.

We now consider the contribution of technical noise to the SNR of a beam deflection measurement. Suppose that there are $N$ photons contributing to the measurement of a deflection of distance $d$. In addition to the Poisson shot noise $\eta_i$, there is technical noise $\xi(t)$ that we model as a white noise process with zero mean and correlation function $\langle \xi(t)\xi(0) \rangle = S^2_\xi \delta(t)$. The measured signal $x = d + \eta_i + \xi(t)$ then has contributions from the signal, the shot noise, and the technical noise. The variance of the time-averaged signal $\bar x$ is given by $\Delta {\bar x}^2 = (1/N^2) \sum_{i,j=1}^N  \langle \eta_i \eta_j \rangle + (1/t^2) \int_0^t dt' dt'' \langle \xi(t') \xi(t'')\rangle$, where the shot noise and technical noise are assumed to be uncorrelated with each other. For a coherent beam described in Eq.\ (\ref{intensity}), the shot noise variance is $\langle \eta_i \eta_j \rangle = \sigma^2 \delta_{ij}$. Therefore, given a photon rate $\Gamma$ (so $N = \Gamma t$), the measured distance
(after integrating for a time $t$) is given by
\begin{equation}
\langle x\rangle = d \pm \frac{\sigma}{\sqrt{\Gamma t}} \pm \frac{S_\xi}{\sqrt{t}}.
\end{equation}

We now compare this with the weak value case.  Given the same number of original photons $N$, we will only have $P_{ps} N$ post-selected photons, while the technical noise stays the same.  Taking $d \rightarrow {\cal A} d$ this gives
\begin{equation}
\langle x\rangle = \frac{1}{\sqrt{P_{ps}}} \left(\alpha  d \pm \frac{\sigma}{\sqrt{ \Gamma t}} \pm \frac{S_\xi \sqrt{P_{ps}}}{\sqrt{t}} \right).
\end{equation}
In other words, once we rescale, we have the same enhancement of the SNR by $\alpha$ as discussed in Eq.\ (\ref{SNRA}), but additionally the technical noise contribution is reduced by $\sqrt{P_{ps}}$ from using the weak value post-selection. Therein lies the power of weak value amplification for reducing the technical noise of a measurement.

The experimental setup is shown in Fig. 1. A 780 nm fiber-coupled laser is launched and collimated using a $20\times$ objective lens followed by a spherical lens with $f = 500$ mm (not shown) to produce a collimated beam radius of $\sigma = 1.7$ mm. For smaller beam radii, the lens is removed and the $20\times$ objective is replaced with a $10\times$ objective. A polarizer is used to produce a pure horizontal linear polarization. The beam enters the interferometer (this is the pre-selection) and is divided, traveling clockwise and counterclockwise, before returning to the beamsplitter (BS). A piezo-actuated mirror on a gimbal mount at a symmetric point in the interferometer is driven (horizontally) with a 10 kHz sine wave with a flat peak of duration 10 $\mu$s. The piezo actuator moves 127 p.m./mV at this frequency with a lever arm of 3.5 cm. Due to a slight vertical misalignment of one of the interferometer mirrors, the output port does not experience total destructive interference (this is the post-selection on a nearly orthogonal state) and contains approximately 20\% of the total input power, corresponding to $\phi/2 = 25^\circ$. A second beamsplitter sends this light to a quadrant cell detector (QCD) (New Focus model 2921) and a charge coupled device (CCD) camera (Newport model LBP-2-USB). The output from the CCD camera is monitored and the output from the quadrant cell detector is fed into two low-noise preamplifiers with frequency filters (Stanford Research Systems model SR560) in series. The first preamplifier is ac coupled with the filter set to 6 dB/oct band-pass between 3 and 30 kHz with no amplification. The second preamplifier is dc coupled with the filter set to 12 dB/oct low-pass at 30 kHz and an amplification factor ranging from 100 to 2000. The low-pass filter limits the laser noise to the 10--90\% risetime of a 30 kHz sine wave ($\tau = 10.5 \mu$s) and so we take this limit as our integration time such that the number of photons incident on the detector is $N = P \tau/E_\gamma$ where $P$ is the power of the laser and $E_\gamma$ is the energy of a single photon at $\lambda = 780$ nm. 

\begin{figure}
\centerline{\includegraphics[width=90mm,height=130mm]{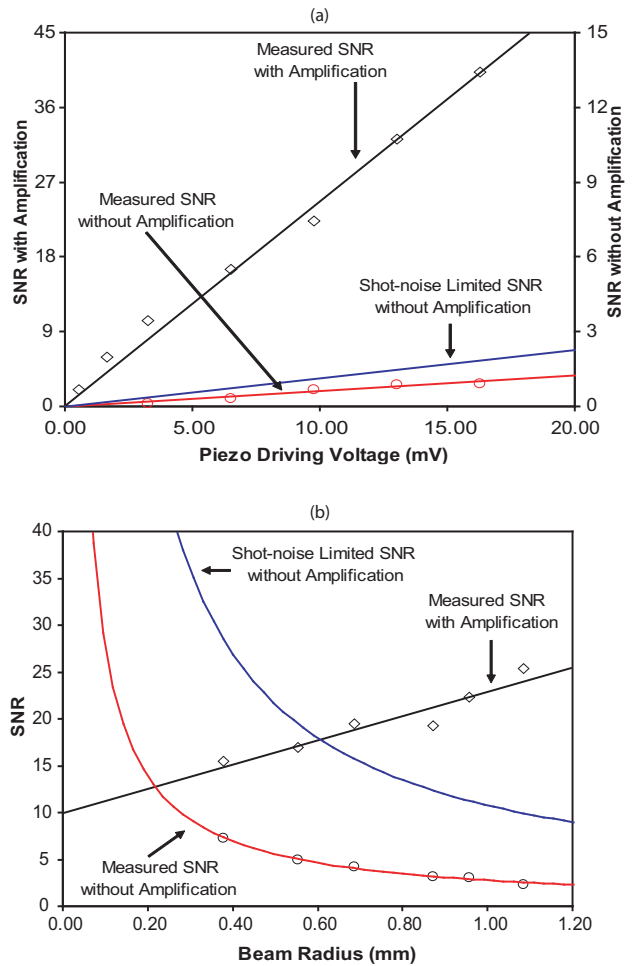}} 
\caption{(Color online) The SNR for \textit{SD setup} (blue curves) is calculated using Eq.\ (\ref{SNR}) assuming perfect quantum efficiency. The SNR was measured with (diamonds, black curves) and without (circles, red curves) the weak value amplification. As predicted by Eq.\ (\ref{SNRA}), (a) shows the dependence on driving voltage (and hence deflection $d$). (b) shows the dependence on beam radius as predicted by Eqs. (\ref{SNR}) and (\ref{SNRA2}). Note that for (a), the black curve is plotted using the left axis whereas the blue and red curves are plotted using the right axis. The lines are linear or $1/\sigma$ fits. The $y$-intercepts of the linear fits in (a) are forced to zero. The statistical variations are smaller than the data points.}
\end{figure}

In what follows, we compare measurements using two separate configurations: \textit{WVA setup} is shown in Fig. 1 and produces the weak value amplification SNR found in Eq.\ (\ref{SNRA}); \textit{SD setup} (for standard detection) is the same as \textit{WVA setup} but with the first 50/50 beamsplitter removed, resulting in the SNR given by Eq.\ (\ref{SNR}). The theoretical curves of the SNR in Fig. 2, to which our data are compared, assume the configuration of \textit{SD setup} with a noiseless detector which has a perfect quantum efficiency; this is what we refer to as an ``ideal measurement." We see reasonable agreement of the data with theory by noting the trends in Fig. 2 as predicted by Eqs. (\ref{SNRA}) and (\ref{SNRA2}). The quoted error comes from the measured data's standard deviation from the linear fits. 

Data were taken for a fixed beam radius $\sigma = 1.7$ mm and detector distance $l_{md} = 14$ cm for two cases: (1) a variable piezo actuator driving voltage amplitude with a fixed input power of 1.32 mW [Fig. 2a]; and with (2) a variable input power with a fixed driving voltage amplitude of 12.8 mV (not graphed). For the first case, using \textit{SD setup}, we measured a SNR a factor of $1.77\pm0.07$ worse than an ideal measurement; with WVA, i.e. \textit{WVA setup}, an improvement of $39\pm3$ was obtained, corresponding to a SNR that is a factor of $21.8\pm0.5$ better than an ideal measurement using \textit{SD setup}. For the second case, we found that the SNR with WVA was linear in power, resulting in a SNR a factor of $22.5\pm0.5$ better than an ideal measurement using \textit{SD setup}. 

Next, the beam radius at the detector $\sigma$ was varied from 0.38 to 1.1 mm while the beam radius at the lens was roughly constant at $a = 850$ $\mu$m. For this measurement, the input power was 1.32 mW, the distances were $l_{lm} = 0.51$ m and $l_{md} = 0.63$ m, and the driving voltage amplitude was 12.8 mV. The results are shown in Fig. 2b. Using \textit{SD setup}, we find that the SNR varies inversely with beam radius as predicted by Eq.\ (\ref{SNR}). However, using \textit{WVA setup}, we see a linear increase in the SNR as the beam radius is increased as predicted by Eq.\ (\ref{SNRA2}).

To verify the dependence of the SNR on $l_{md}$, as seen in Eqs. (\ref{SNR}) and (\ref{SNRA}), we fixed the input power at 1.32 mW, the driving voltage amplitude at 12.8 mV, the beam radius at $\sigma = 1.7$ mm and varied the position of the detector relative to the piezo-actuated mirror. We found that, using \textit{WVA setup}, the SNR was roughly constant with a value of 29 $\pm1$. This can be understood by realizing that, in Eq.\ (\ref{SNRA}), the $l_{md}$ in the denominator cancels the $l_{md}$ in the numerator owing to the fact that $d = l_{md} (\Delta\theta)$, where $\Delta\theta$ is the angular deflection. Using \textit{SD setup}, we saw the expected linear relationship and we found that the system is worse than an ideal system by a factor of $3.2\pm0.1$.

To demonstrate the utility of this method we constructed a smaller interferometer with a smaller $l_{md} = 42$  mm and a smaller beam radius $\sigma = 850$ $\mu$m. For this geometry with 2.9 mW of input light and 390 $\mu$W of output light, the predicted amplification $\alpha = 260$. With these parameters, the SNR for an ideal \textit{WVA setup} is approximately unity. We measured $\alpha$ to be 150. Combining this with our nonideal detector, we obtain an improvement of the SNR better than a quantum limited \textit{SD setup} by a factor of 54. Practically, this means that in order to obtain equal measurement precision with this quantum limited system using the same beam radius it would take over three more orders of magnitude of time or power. 

An important note is that the expected WVA of the SNR for the larger interferometer is approximately $\alpha = 300$; yet only an $\alpha = 55$ (a factor of 5.5 below) was obtained from the graphed data. However, for the smaller interferometer, the measured $\alpha$ was only a factor 1.7 below the predicted value. 

The connection between standard deflection measurement techniques and the weak value scheme presented here will be elucidated at a later time. While this method does not beat the ultimate limit for a beam deflection measurement, it does have a number of improvements over other schemes: (1) the reduction in technical noise; (2) the ability to use high power lasers with low power detectors while maintaining the optimal SNR; and (3) the ability to obtain the ultimate limit in deflection measurement with a large beam radius. Additionally, we point out that, while weak values can be understood semi-classically, the SNR in a deflection measurement requires a quantum mechanical understanding of the laser and its fluctuations. 

It is interesting to note that interferometry and split detection have been competing technologies in measuring a beam deflection \cite{Putman1992}. Here we show that the combination of the two technologies leads to an improvement that can not be observed using only one, i.e.\ that measurements of the position of a large radius laser beam with WVA allows for better precision than with a quantum limited system using split detection for the same beam radius. Applications that can take advantage of this setup include: measuring the surface of an object by replacing the piezo actuator with a stylus such as with atomic force microscopy; or measuring frequency changes due to a dispersive material such as in Doppler anemometry. 

This work was supported by DARPA DSO Slow Light, a DOD PECASE award, and the University of Rochester.


\end{document}